\documentclass[twocolumn,aps,prl,showpacs]{revtex4}
\usepackage{graphics}
\usepackage{amsmath}
\def\beq{\begin{equation}}
\def\eeq{\end{equation}}
\def\bea{\begin{eqnarray}}
\def\eea{\end{eqnarray}}

\def\nnu{\nonumber}
\def\tst{\textstyle}
\def\dst{\displaystyle}

\def\eno#1{Eq.~(\ref{#1})}

\def\al{\alpha}
\def\be{\beta}

\def\eps{\epsilon}
\def\tta{\theta}

\def\vph{\varphi}

\def\Om{\Omega}

\def\by{\over}
\def\grad{\boldsymbol{\nabla}}

\def\hf{{1\over2}}

\def\tshf{\tst\hf}

\def\tofro{\leftrightarrow}

\def\gtwid{\mathrel{\raise.3ex\hbox{$>$\kern-.75em\lower1ex\hbox{$\sim$}}}}
\def\ltwid{\mathrel{\raise.3ex\hbox{$<$\kern-.75em\lower1ex\hbox{$\sim$}}}}

\def\ham{{\cal H}}
\def\ket#1{|#1\rangle}

\def\olap#1#2{\langle#1|#2\rangle}
\def\mel#1#2#3{\langle#1|#2|#3\rangle}

\def\itx{{\it x\/}}
\def\itxy{{\it xy\/}}
\def\itxz{{\it xz\/}}

\def\bB{{\bf B}}
\def\bH{{\bf H}}
\def\bJ{{\bf J}}
\def\bh{{\bf h}}
\def\br{{\bf r}}

\def\bu{{\bf u}}
\def\bs{{\bf s}}
\def\bv{{\bf v}}
\def\bw{{\bf w}}
\def\bg{{\bf g}}

\def\bt{{\bf t}}

\def\xhat{{\bf{\hat x}}}
\def\yhat{{\bf{\hat y}}}
\def\zhat{{\bf{\hat z}}}

\def\Fe8{Fe$_8$}

\begin{document}


\title{Lost and found: The missing diabolical points in the Fe$_8$ 
molecular magnet}

\author{Feifei Li}
\author{Anupam Garg}
\email[e-mail address: ]{agarg@northwestern.edu}
\affiliation{Department of Physics and Astronomy, Northwestern University,
Evanston, Illinois 60208}

\date{\today}

\begin{abstract}
Certain diabolical points in the tunneling spectrum of the single-molecule
magnet \Fe8 were previously believed to be have been eliminated as a result
of a weak fourth-order anisotropy. As shown by Bruno, this is not so, and
the points are only displaced in the magnetic field space along the medium
anisotropy direction. The previously missing points are numerically located
by following the lines of the Berry curvature. The importance of an
experimental search for these rediscovered points is discussed.
\end{abstract}


\maketitle

The purpose of this note is to report on a numerical search of certain
diabolical points (DP's) in the energy spectrum of the single-molecule
magnet (also known as a molecular magnet or nanomagnet) \Fe8 that were earlier
believed to be missing, but are in fact not so~\cite{bruno}.
Several other DP's {\it have\/} been seen experimentally
in \Fe8~\cite{werns}, and their observation provides the best evidence
of spin orientation tunneling between deep levels in all single-molecule
magnets studied to date. Observation of even some of the missing DP's would
strengthen our understanding of \Fe8 substantially.

For a system whose Hamiltonian depends on some parameter, a DP is a
point in parameter space where two (or more) energy levels are
degenerate~\cite{bw84}.
In \Fe8, the parameter is the static applied magnetic field, and the
locations of the DP's so far observed (as well as many other experimental
measurements) are well described by the anisotropy Hamiltonian,
\beq
 \ham = k_1 J_x^2  + k_2 J_y^2 - C(J_+^4 + J_-^4)
         - g \mu_B \bJ\cdot\bH.
 \label{ham}
\eeq 
Here $\bJ = (J_x, J_y, J_z)$ is the spin, $\bH$ is an
external magnetic field, $g$ is a g-factor, $\mu_B$ is the Bohr magneton, 
and $k_1$, $k_2$, and $C$ are anisotropy coefficients. Experimentally,
$J =10$, $k_1 \simeq 0.338$~K, $k_2\simeq 0.246$~K,
$C \simeq 29\,\mu$K, and $g\simeq 2$. The DP's can be understood as
arising when tunneling between two states with (nearly) oppositely
oriented magnetic moment is quenched because of destructive interference
between instantons (spin tunneling
trajectories)~\cite{gargepl,loss_and_henley}.

The model Hamiltonian (\ref{ham}) was first analyzed in 1993~\cite{gargepl}
with $C=0$, and it was found that for ground state tunneling it had 10
DP's along the positive $H_x$ axis, corresponding to $J=10$. In reality
only 4 DP's are seen~\cite{werns}, which was explained in
Ref.~\cite{ekag} as follows. When $C\ne 0$, we get two new (but
noninterfering) instantons, which are discontinuous at the end points.
One of these instantons has the least action when $H_x$ exceeds a certain
value $H^*$, and since this instanton
has no interfering partner, there are no more DP's for $H_x > H^*$.
For the values of the anisotropy coefficients quoted above,
$H^*$ lies just beyond the location of the fourth DP, which explains
why the last six DP's are not seen in experiment or by direct numerical
diagnolization of \eno{ham}. We can further check this picture by
decreasing $C$ so as to make $H^*$ larger. It is seen that as this
happens, one successively sees six and then eight, and finally all ten
DP's, all in accord with direct numerical diagonalization
results~\cite{ekag}.

However, as shown by Bruno~\cite{bruno}, the above picture, though
correct, is incomplete. For any energy level, the sum of the Chern
numbers for all DPs involving that level is a topological invariant as
parameters like $k_1$, $k_2$, or $C$ are varied. Since DP's in any
system are generically simple, we expect this to be so in \Fe8 also,
and the Chern number for any one DP should be $\pm 1$ whether
$C=0$ or $C\ne 0$. Hence the six missing DP's must be present
elsewhere in magnetic field space. A similar conclusion applies to
the DP's associated with tunneling between other pairs of
levels~\cite{ekag01}. For tunneling between the ground states, the
DP's merely move off the \itx\ axis into the \itxy\ plane.
For the higher energy levels, they  move off the \itxz\ plane into the
full three dimensional $\bH$ space. This point can also be understood
by noting that for a system with purely four-fold
symmetry ($k_2 = k_1$, $C \ne 0$), the ground state DP's lie on the
$\pm\xhat \pm \yhat$ axes, while for the excited states they lie in the
planes formed by these axes and the $\zhat$ axis~\cite{cpag}. When both
two-fold and four-fold anisotropies are present ($k_2 \ne k_1$,
$C \ne 0$), it is then not surprising that the location of some of the
DP's is also intermediate~\cite{bruno}.

Observation of these rediscovered DPs would be interesting in itself,
and also provide an important test of the validity of the model
(\ref{ham}) vs. other models~\cite{leuen} that add extra 6th and
8th order anisotropies because the location of the DPs is very
sensitive to the higher order anisotropies. With this motivation, we
have undertaken a search for the DP's for the ground state and some of
the excited states. We stress that the key insight that these points
should exist in the first place is due to Bruno, and our contribution
is only to find their specific locations. Neverthless, finding them
is not without challenge as we discuss next.

\begin{figure}
\includegraphics{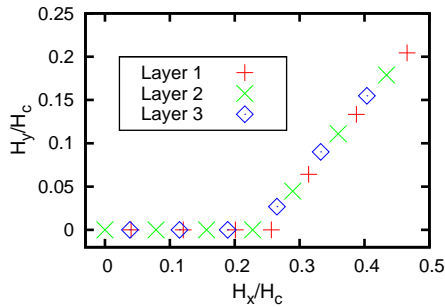}
\caption{\label{layers}  Diabolical points of \Fe8 corresponding to
tunneling from the Zeeman level $m = -10$ to $m =10$ (layer 1), to
$m=9$ (layer 2), and to $m=8$ (layer 3), all projected onto the \itxy\ plane.
$H_c = 2k_1 J/g\mu_B$.}
\end{figure}

A direct search for the DPs by numerical minimization of the
energy differences fails because the energy surface is like a golf
course with rolling hills on which the DPs are the holes. Because
the holes are so localized, unless one starts close to one of them
by luck, any numerical algorithm will in general simply head
for the valleys
of the course and miss the holes entirely. Because $\ham$ is not
real for general $\bH$, we also cannot use the method of
Ref.~\cite{cpag}, which is to corral the DPs by using the Herzberg and
Longuet-Higgins theorem~\cite{hlh63} to find and successively bisect a
sign-reversing circuit. We therefore proceed as
follows. Let us denote the eigenstates and eigenvalues of \eno{ham}
for fixed $\bH$ by $\ket{n(\bH)}$ and $E_n(\bH)$,
$n = 1,\ 2,\ldots, 21$, and order them so that $E_n \ge E_{n-1}$
for every $\bH$. Except at degeneracies (the DPs), the Berry
curvature for the $n$th level is defined by \cite{berry}
\beq
\bB_n
   = -{\rm Im} \sum_{n'\ne n}
       {\mel{n}{\grad_{\bH}\ham}{n'} \times
             \mel{n'}{\grad_{\bH}\ham}{n} \by (E_{n'} - E_n)^2}.
                             \label{B_curv2}
\eeq
\begin{table}
\caption{\label{dplocs}Locations of selected diabolical points}
\begin{ruledtabular}
\begin{tabular}{l c l}
$(m,m')$ & Layer & $(h_x, h_y, h_z)$ \\
\hline
$(-10,10)$ & 1 & $(0.0404,0,0)$ \\
&& $(0.1207, 0, 0)$ \\
&& $(0.2011, 0, 0)$ \\
&& $(0.2565, 0, 0)$ \\
&& $(0.3139, \pm0.0642, 0)$ \\
&& $(0.3875, \pm0.1334, 0)$ \\
&& $(0.4656, \pm0.2043, 0)$ \\
\hline
$(-10,9)$ & 2 & $(0,0,0.0426)$ \\
&& $(0.0787,0,0.0426)$ \\
&& $(0.1568,0,0.0427)$ \\
&& $(0.2276,0,0.0427)$ \\
&& $(0.2893,\pm0.0447,0.0427)$ \\
&& $(0.3594,\pm0.1109,0.0427)$ \\
&& $(0.4340,\pm0.1788,0.0427)$ \\
\hline
$(-10,8)$ & 3 & $(0.0385,0,0.0853)$ \\
&& $(0.1150,0,0.0853)$ \\
&& $(0.1892,0,0.0853)$ \\
&& $(0.2650,\pm0.0268,0.0854)$ \\
&& $(0.3326,\pm0.0900,0.0854)$ \\
&& $(0.4038,\pm0.1548,0.0854)$ \\
\hline
$(-9,9)$ & 1 & $(0.0384, 0, 0)$ \\
&& $(0.1147, 0, 0)$ \\
&& $(0.1889, 0, 0)$ \\
&& $(0.2651, \pm0.0261, 0)$ \\
&& $(0.3325, \pm0.0891, 0)$ \\
&& $(0.4034, \pm0.1535, 0)$ \\
\hline
$(-8,8)$ & 1 & $(0.0365, 0, 0)$ \\
&& $(0.1089, 0, 0)$ \\
&& $(0.1805, 0, 0)$ \\
&& $(0.2339, 0, 0)$ \\
&& $(0.2831, \pm0.0511, 0)$ \\
&& $(0.3468, \pm0.1092, 0)$ \\
\end{tabular}
\end{ruledtabular}
\end{table}
Further, the Chern number associated with a degeneracy is given by
\beq
Q_n = -{1\by 2\pi} \oint_S \bB_n \cdot d^2\bs,
                 \label{Chern}
\eeq
where the integral is independent of the choice of the surface $S$,
as long as it encloses the degeneracy, since
$\grad_{\bH}\cdot\bB_n = 0$ away from the degeneracy.
The Chern number is always an integer, and for a simple
double degeneracy, it equals $\pm 1$. In other words, near a DP, $\bB_n$
has the form of a monopole field with flux equal to $\pm 2\pi$.
Hence, to find the DPs, we numerically evaluate $\bB_n$ for an
initial $\bH$, and follow the lines of $\bB_n$ in the direction of
increasing strength until we hit a monopole. Since the number
of DPs where levels $n$ and $n+1$ are degenerate is topologically fixed
and known, all the DPs can be found by taking sufficiently many initial
values of $\bH$. The DPs for successive pairs of levels occur in layers,
with $H_z$ essentially constant in a layer.
With $\bh \equiv g\mu_B \bH/2k_1 J$, the first three layers are at
$h_z = 0$ (exactly), $0.0427$, and $0.0854$. It should be noted, however,
that in a given layer, one can have DP's corresponding to tunneling
between levels with different pairs of Zeeman quantum numbers. For example,
layer 1 contains points corresponding to tunneling between
$m = -10 \tofro 10$, $-9 \tofro 9$, $-8 \tofro 8$, etc.
Likewise, layer 2 contains points corresponding to tunneling between
$m = -10 \tofro 9$, $-9 \tofro 8$, $-8 \tofro 7$, etc. In
Table~\ref{dplocs}, we show the DP's for $(-10\tofro 10)$
$(-10\tofro 9)$, $(-10\tofro 8)$, $(-9 \tofro 9)$, and $(-8 \tofro 8)$
tunneling. For the first three of these pairs of levels, the projections
of the DP's onto the \itxy\ plane are shown in Fig.~\ref{layers}.
We note that it is just these three pairs of levels for which tunnel
splittings were reported in Ref.~\cite{werns}. Hoewever, except for the
ground pair of levels (layer 1), not even all the DP's on the \itx\ axis
are found there.

In the rest of this note, we discuss the form of the Berry curvature
near a DP in more detail. For simplicity, we will divide $\ham$ by
$k_1 J^2$. Since $\grad_{\bH}\ham$ and $E_n$ are both divided by this
factor, it follows from \eno{B_curv2} that $\bB_n$ is unchanged. With this
preliminary remark, let us suppose that $E_n = E_{n-1} \equiv E_{n,n-1}$
at $\bh = \bh_{n,n-1}$, and denote
\beq
\br = \bh - \bh_{n,n-1}.
\eeq
Further, let us make a particular choice of the two degenerate states
at $\br = 0$, and denote them by $\ket{a}$ and $\ket{b}$, with
$\olap{b}{a} = 0$. (Any orthogonal linear combination of $\ket{a}$
and $\ket{b}$ would also work.) It suffices to truncate the Hamiltonian
to this two dimensional subspace since the sum in \eno{B_curv2} is
dominated by degenerate states. Hence, at $\br = 0$, we have
\beq
\ham = E_{n,n-1} \begin{pmatrix}
       1 & 0 \\
       0 & 1 \\
                 \end{pmatrix}.
\eeq
For small enough $\br$, we can take $\ket{a}$ and $\ket{b}$ to be
unchanged, so
\beq
\grad_{\bh}\ham = -{2\by J}
%
  \begin{pmatrix}
     \bJ_{aa}  &  \bJ_{ab} \\
     \bJ_{ba}  &  \bJ_{bb}
  \end{pmatrix},
\eeq
where $\bJ_{aa} = \mel{a}{\bJ}{a}$ etc. Next, let us define
$\bJ_{aa} + \bJ_{bb} = J\bt$,
$\bJ_{aa} - \bJ_{bb} = J\bu$,
$\bJ_{ab} = J(\bv + i\bw)/2$, where $\bt$, $\bu$, $\bv$, and $\bw$ are
real vectors. In terms of these vectors, we have
\beq
\ham = - {\bt\cdot\br \by J} - 
 \begin{pmatrix}
   \bu\cdot\br  &   (\bv + i\bw)\cdot\br \\
   (\bv - i\bw)\cdot\br  & -\bu\cdot\br
 \end{pmatrix},
\eeq
where we have ignored the constant $E_{n,n-1}$. Similarly ignoring the
overall shift $-\bt\cdot\br/J$, the eigenvalues of this matrix
are $\pm\eps(\br)$, with
\beq
\eps(\br) = [(\bu\cdot\br)^2 + (\bv\cdot\br)^2 + (\bw\cdot\br)^2]^{1/2}.
\eeq
To write the eigenvectors compactly, we define
\bea
\cos\tta(\br) &=& \bu\cdot\br/\eps(\br), \\
\sin\tta(\br) e^{i\vph(\br)}
              &=& (\bv + i \bw)\cdot\br/\eps(\br).
\eea
The eigenvectors are then
\beq
\ket{-} = 
  \begin{pmatrix}
    \cos\tshf\tta \\ \sin\tshf\tta e^{-i\vph}
  \end{pmatrix},
\quad
\ket{+} = 
  \begin{pmatrix}
    \sin\tshf\tta \\ -\cos\tshf\tta e^{-i\vph}
  \end{pmatrix}.
\eeq
Further abbreviating $c = \cos\hf\tta$ and $s = \sin\hf\tta$, and
$\bg = \mel{+}{\grad_{\bh}\ham}{-}$, we have
\beq
\bg = -2cs\bu - s^2 e^{-i\vph} (\bv + i\bw)
        + c^2 e^{i\vph} (\bv - i \bw).
\eeq
It then follows that
\bea
\bg \times \bg^* 
 = 2i [\cos\tta (\bv\times\bw)
          &+& \sin\tta \cos\vph (\bw\times\bu) {\hskip1cm} \nnu\\
           &{}& \!\!\!\! {} + \sin\tta \sin\vph (\bu\times\bv)],
\eea
so that for the level labeled ${+}$, 
\bea
\bB_{+}
   &=& {-1\by 2\eps^3(\br)}
      \left[ (\bu\cdot\br) (\bv\times\bw)
                 + (\bv\cdot\br) (\bw\times\bu) \right. \nnu \\
   &&{\hskip3.3cm}  \left.{} + (\bw\cdot\br) (\bu\times\bv) \right]  \nnu \\
   &=& {-1\by 2\eps^3(\br)}
      \left[ \bu\cdot(\bv\times\bw) \right] \br.
\eea
This is clearly of monopole form with appropriately scaled and sheared
axes. It is not difficult to show that $\bB_{-} = -\bB_{+}$ and that
$\grad\cdot\bB_{+} = 0$.

To find the Chern number, we must evaluate the integral
\beq
Q_{\pm} = -{1\by 2\pi}\oint_S \bB_{\pm}\cdot d^2\bs
\eeq
for a suitable surface $S$. Let us take $S$ to be the surface of
 the parallepiped with vertices at $(\pm\bu^*, \pm\bv^*, \pm\bw^*)$,
where $\bu^*$, $\bv^*$, and $\bw^*$ are the reciprocal vectors
\bea
&\bu^* = {\dst{\bv \times \bw \by\Om}} , \quad
 \bv^* = {\dst{\bw \times \bu \by\Om}}, \quad
 \bw^* = {\dst{\bu \times \bv \by\Om}};& \\
&\Om = \bu\cdot (\bv\times \bw).&
\eea
Then $\bu\cdot \bu^* =1$, $\bu\cdot\bv^* = \bu\cdot\bw^* = 0$, etc.
Let us consider the integral over the face of the parallepiped that
has edges along $\bv^*$ and $\bw^*$, and thus
has a normal parallel to $+\bu$ (the others have normals along
$-\bu$, $\pm\bv$, and $\pm\bw$). On this face, we may parametrize
$\br$ as
\beq
\br = \bu^* + \al \bv^* + \be \bw^*,
\eeq
where $-1 \le \al \le 1$, $-1 \le \be \le 1$. Then, since the area
of this face is $4|\bv^*\times\bw^*| = 4|\bu|/\Om$,
\beq
d^2\bs = {\bu \by \Om} d\al\, d\be.
\eeq
Finally, $\bu\cdot\br =1$, $\bv\cdot\br =\al$, and $\bw\cdot\br = \be$.
Thus, the contribution from this face to $Q_{\pm}$ is given by
\beq
\pm {1\by 4\pi}
  \int_{-1}^1 \int_{-1}^1 {d\al\, d\be \by (1 + \al^2 + \be^2)^{3/2}}
       = \pm{1\by 6}.
\eeq
(The integral is elementary, and can be performed by standard
trigonometric substitutions.) The contributions from the other faces
are identical, so
\beq
Q_{\pm} = \pm 1.
\eeq

Thus the Chern number associated with the lower energy level is always
$-1$ irrespective of the details, and this is why the sum of the numbers
for a given energy level cannot be altered by varying the anisotropy
parameters.

The reader cannot have failed to notice the regularity of the DP pattern
for the present problem. This is in sharp contrast to the generic
situation where no pattern is expected apart from a general scaling of the
density with energy~\cite{bw84}. The reason for this regularity is unclear
to us.

This work is supported by the NSF via grant no. PHY-0854896.


%
\end{document}